# Pressure-Enhanced Amplified Spontaneous Emission in Solvent-Engineered $MAPbBr_3$ Films

Xin Tang, Linrui Li, Shuaiqi Li,* and Dingke Zhang*

College of Physics and Electronic Engineering, Chongqing Normal University
400000 Chongqing (China)
E-mail: Copper@cqnu.edu.cn (Xin Tang), Copper@cqnu.edu.cn (Shuaiqi Li)

**Abstract:** The high-pressure response of three-dimensional metal halide perovskites (MHPs) has been extensively explored, with prior studies focusing predominantly on pressure-induced structural transitions and photoluminescence behavior. However, the evolution of their laser-related optical gain under compression remains largely uninvestigated. Herein, we employ solvent engineering to fabricate $MAPbBr_3$ thin films using the ionic liquid methylammonium acetate (MAAc) as a benign processing medium ($MAPbBr_3$-MAAc), and compare them with reference films prepared from conventional N,N-dimethylformamide (DMF) and dimethyl sulfoxide (DMSO) solvents ($MAPbBr_3$-DMF:DMSO). Through *in situ.* high-pressure optical spectroscopy, we systematically characterize the amplified spontaneous emission (ASE) properties alongside photoluminescence and absorption. Complementary high-pressure Raman spectroscopy reveals that the optical changes are directly correlated with pressure-driven structural phase transitions. Importantly, the $MAPbBr_3$-MAAc films exhibit all delayed transition pressure points and superior pressure resilience: they retain ASE emission up to 3.98 GPa, significantly higher than the 3.34 GPa limit of the $MAPbBr_3$-DMF:DMSO films, and demonstrate consistently lower ASE thresholds and longer carrier lifetimes across the studied pressure range. This work not only advances the understanding of high-pressure optoelectronic behavior in MHPs, but establishes solvent engineering as an effective strategy for designing robust perovskite gain media capable of functioning in extreme environments.

## 1. Introduction

Metal halide perovskites (MHPs) have rapidly established themselves as highly promising semiconductor materials for optoelectronic applications, owing to their exceptional optoelectronic properties such as tunable direct bandgaps, high photoluminescence quantum yields, long carrier diffusion lengths, low defect densities, and facile solution processability [1-5]. These advantages have facilitated the demonstration of optically pumped lasing in various MHP configurations, including thin films, nanowires, and microcavity structures, with notably low thresholds and high-quality factors, underscoring the potential of perovskites for developing compact and low-cost coherent light sources [6-8]. For instance, Murzin et al. reported amplified spontaneous emission (ASE) and random lasing in $MAPbBr_3$ single crystals, further affirming the material's suitability for laser applications [9]. Weng et al. also achieved single-mode lasing in $MAPbBr_3$ microcavities across an ultra-broad pump wavelength range of 400-2300 nm, with the lasing output remaining stable under high-power excitation for over 12 hours [10]. More recently, an electrically driven perovskite laser based on a dual-cavity architecture was proposed, exhibiting a low lasing threshold and highlighting a feasible pathway toward integrated optoelectronic devices [11]. Currently, research efforts in MHPs for laser applications have predominantly focused on optimizing their performance under ambient conditions. This is achieved mainly by reducing ASE thresholds and enhancing lasing efficiency through approaches such as morphology control, defect passivation, and waveguide engineering [12-14]. Techniques including surface passivation, polymer encapsulation, and additive engineering have been successfully implemented to suppress non-radiative recombination, improve crystalline quality, and boost ASE efficiency in perovskite films and nanocrystals [15-17].

However, as perovskite optoelectronic devices-including MHP-based lasers-are increasingly considered for extreme environments (e.g., spaceborne photonics, underwater optical communication, and harsh-environment sensing), the demand for environmentally robust perovskite materials has become more pressing [18-20]. In such scenarios, devices may encounter multiple stressors, including mechanical strain, radiation, and high pressure [21]. Among these, pressure stands out as a fundamental and widely applicable tuning parameter that can mimic various extreme conditions. Under high pressure, key material properties, such as crystal structure, band alignment, charge dynamics, and light-matter interactions, can be dramatically altered through compression-induced bandgap shifts and phase transitions [22,23].

Previous studies have shown that pressure can trigger structural phase transitions and markedly modify the optical response of perovskites. For example, $MAPbI_3$ undergoes a pressure-driven transition from a cubic to a tetragonal phase, which significantly changes its optical properties [24]. Its bandgap shifts from direct to indirect, while both optical absorption and photoluminescence are substantially enhanced under compression. Similarly, in $CsPbBr_3$ nanocube superlattices (NC-SLs), pressure-induced phase transitions from cubic to orthorhombic and eventually to an amorphous-like state have been observed, leading to enhanced photoluminescence via surface reconstruction [25]. Despite these insights, systematic investigations into the laser-related optical gain and ASE behavior of perovskites under high pressure remain scarce.

In parallel, solvent engineering has emerged as a central strategy for improving perovskite optical gain and device performance under ambient conditions. The use of binary solvent mixtures, antisolvent treatments, and tailored additives has proven effective in controlling film morphology and crystallinity, resulting in significantly lower ASE thresholds and higher-quality laser output [26]. Notably, a recent study by Jiang et al. systematically demonstrated that solvent engineering can critically regulate perovskite crystallization kinetics and film morphology [27]. This approach offers a promising pathway for enhancing the high-pressure robustness of MHPs by deliberately optimizing their crystallization behavior from a solvent-engineering perspective.

Here, we employed the ionic liquid methylammonium acetate (MAAc) to optimize the crystallization of perovskite thin films. The high-pressure performance of these MAAc-processed films was systematically compared with that of films fabricated using the conventional solvent system N,N-dimethylformamide (DMF) and dimethyl sulfoxide (DMSO). This study aims to investigate the feasibility of enhancing the environmental robustness of MHPs via solvent engineering and to elucidate the intrinsic connection between their laser-related optical gain and ASE behavior under high pressure. Our results demonstrate that $MAPbBr_3$ films prepared from different solvents all undergo a sequential phase transition under pressure ($Pm\overline{3}m \rightarrow Im\overline{3} \rightarrow Pmn2_1$). Notably, the $MAPbBr_3$-MAAc film exhibits significantly enhanced pressure stability, consistently outperforming the $MAPbBr_3$-DMF:DMSO film in ASE performance across the applied pressure range. Furthermore, we employed in situ transient absorption (TA) spectroscopy to probe the carrier dynamics under high-pressure conditions. Through a correlated high-pressure multispectral analysis, this work provides deeper insights into how solvent engineering governs the optical properties and phase-structure stability of MHPs, offering a strategic pathway for optimizing their laser performance in extreme environments.

## 2. Experimental Details

### 2.1 Materials：

Ammonium bromide (MABr) and lead bromide ($PbBr_2$) were purchased from Xi'an Polymer Light Technology Corp. DMSO and DMF were purchased from Chron Chemicals. All the reagents were used directly without further purification.

### 2.2 Preparation of Samples：

$MAPbBr_3$-MAAc film: a clarified $MAPbBr_3$ precursor solution was prepared by mixing $PbBr_2$ (153.25 mg) and MABr (46.75 mg) (molar ratio of $PbBr_2$:MABr is 1:1) in MAAc (1 mL) and stirring for 8 hours at 60 ℃. Then, 150 μL of the $MAPbBr_3$ precursor solution with a concentration of 200 mg/mL was dropped onto the PET substrate (pretreated with UV-ozone for 30 min). The substrate was subsequently rotated at 5000 rpm for 30 s at a constant temperature of 70 °C, followed by annealing at 100 °C for 5 min.

$MAPbBr_3$-DMF:DMSO film: for comparison, a clarified solution was prepared by mixing $PbBr_2$ (306.5 mg) and MABr (93.5 mg) (molar ratio of $PbBr_2$:MABr is 1:1) in 1 mL of mixed solvent of DMF and DMSO (3:1 volume ratio) and stirring for 8 hours at room temperature. 150 μL of $MAPbBr_3$ precursor solution at a concentration of 400 mg/mL was spin-coated onto the PET substrate at 2000 rpm for 20 s. Then, an additional 40 μL of chlorobenzene as an anti-solvent was rapidly dropped onto the film during the final 10 s of spin-coating. Finally, the film was annealed at 100 ℃ for 5 min.

### 2.3 High-Pressure Experiments

The high-pressure experiments were carried out using a diamond anvil cell (DAC) equipped with diamonds of 800 μm culet size. A T301 stainless-steel sheet was selected as the gasket, which was already pre-pressurized to a thickness of 120 μm. Subsequently, a sample chamber with a diameter of 400 μm was laser-drilled at the center of the pre-pressurized gasket. Dimethylsilicone oil (viscosity: 1000 ± 80 mPa $s^{-1}$, Macklin) was employed as the pressure-transmitting medium, and the cell pressure was calibrated using a single-crystal ruby ball.

### 2.4 Measuring Instruments:

The *in suit.* high-pressure Raman spectra were recorded on an integrated spectrometer (LabRAM HR Evolution, Horiba, Japan) with a 633 nm laser serving as the excitation source. In *in suit.* high-pressure steady-state photoluminescence (PL) and absorption spectra measurement, a modified microspectroscopy system (Gora-UVN-FL, Ideaoptics, China) which integrated with a 405 nm excitation laser and a halogen light source was employed. *In suit.* high-pressure ASE spectra were acquired by pumping the films with a Q-switched solid-state Nd:YAG laser (Minilite II, Continuum, USA), which emits 3-7 ns pulses at a repetition frequency of 10 Hz and an excitation wavelength of 355 nm. *In suit.* high-pressure femtosecond transient absorption (fs-TA) measurements were performed using a regeneratively amplified Ti:sapphire laser system (Legend Elite HE+USP-1K-III, Coherent, USA), which generates both the pump and probe beams. The 355 nm pump pulses were derived by passing a fraction of the fundamental laser beam through an optical parametric amplifier. A broadband probe pulse (400-800 nm) was generated by focusing an 800 nm laser pulse into a $CaF_2$ crystal. The temporal delay between the pump and probe pulses was controlled by a motorized delay stage. The two beams were spatially overlapped on the sample enclosed in the DAC, and the transmission changes of the probe beam after passing through the sample were detected using a fiber spectrometer (AvaSpec-ULS2048CL-EVO, Avantes, Netherlands).

## 3. Results and Discussion

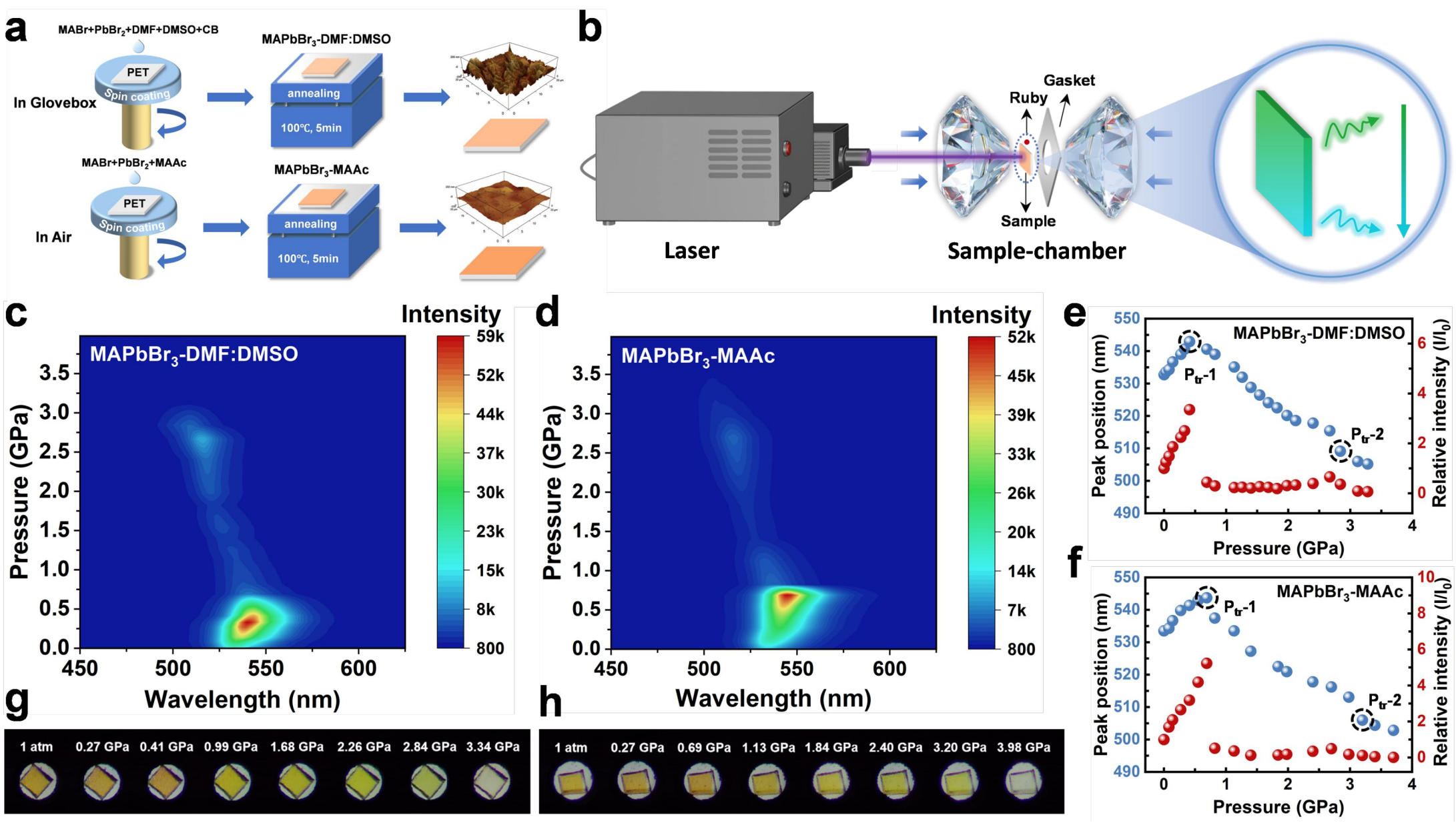

**Figure 1. Film fabrication, high-pressure optical setup, and pressure-dependent PL of $MAPbBr_3$ films.** (a) Schematic of the $MAPbBr_3$ film preparation process. (b) Diagram of the *in situ.* high-pressure optical spectroscopy system using a DAC, with excitation sources for PL and ASE measurements indicated. Three-dimensional plots of PL intensity versus emission wavelength and pressure for the (c) $MAPbBr_3$-DMF:DMSO and (d) $MAPbBr_3$-MAAc films. Evolution of the PL peak position (solid line) and normalized integrated intensity (dashed line) as a function of pressure for the (e) $MAPbBr_3$-DMF:DMSO and (f) $MAPbBr_3$-MAAc films. Optical micrographs of the (g) $MAPbBr_3$-DMF:DMSO and (h) $MAPbBr_3$-MAAc films inside the DAC at selected pressures during compression.

At ambient pressure, $MAPbBr_3$ crystallizes in a cubic perovskite structure (space group $Pm\overline{3}m$) with a lattice constant of approximately 5.9 Å, comprising corner-sharing $[PbBr_6]^{4-}$ octahedra and A-site methylammonium cations ($MA^+$) [28]. This structure gives rise to characteristic green PL. The as-prepared $MAPbBr_3$ films exhibit an orange-yellow appearance. A schematic of the fabrication process is presented in **Figure 1a**. In this work, $MAPbBr_3$ films were prepared on flexible PET substrates. The optimized films were fabricated in air using a one-step thermal spin-coating method with the ionic liquid MAAc as solvent, denoted as $MAPbBr_3$-MAAc. For comparison, reference films were synthesized in a glovebox via the antisolvent method using a conventional DMF:DMSO mixed solvent system, denoted as $MAPbBr_3$-DMF:DMSO. As illustrated in **Figure 1b**, the *in suit.* measurement system employed a continuous-wave laser for PL excitation and a nanosecond pulsed laser for ASE measurements, respectively.

The influence of pressure on the optical properties of the solvent-engineered films was first investigated by *in suit.* PL spectroscopy. At ambient pressure, the $MAPbBr_3$-DMF:DMSO film shows a band-edge PL peak at 532 nm with a full width at half maximum (FWHM) of 20.7 nm, while the $MAPbBr_3$-MAAc film exhibits a slightly red-shifted peak at 534 nm with a narrower FWHM of 19.8 nm (**Figure S1**). The pressure-dependent evolution of the normalized PL spectra is shown in **Figure S2**, with the integrated intensity trends displayed in **Figure 1c** and **1d**.

For the $MAPbBr_3$-DMF:DMSO film (**Figure 1c**), the PL intensity increases sharply upon compression, reaching a maximum at 0.41 GPa ($P_{tr}$-$1$), before gradually decreasing and nearly vanishing at 3.34 GPa. In contrast, the $MAPbBr_3$-MAAc film (**Figure 1d**) demonstrates superior pressure stability: its PL intensity peaks at a higher $P_{tr}$-$1$ of 0.69 GPa and remains detectable up to 3.98 GPa. The emission wavelengths of both films initially show a red-shift of about 10 nm at $P_{tr}$-$1$, followed by a blue-shift until PL quenching (**Figures 1e and 1f**). Notably, the maximum PL intensity coincides with the peak red-shift to ~545 nm below 1 GPa, a regime associated with a reported[29] isostructural phase transition from $Pm\overline{3}m$ to $Im\overline{3}$. Within this range, the PL intensity of the $MAPbBr_3$-DMF:DMSO film is enhanced by approximately 3.4 times relative to its ambient value, while the $MAPbBr_3$-MAAc film exhibits a more pronounced enhancement of about 5.3 times. Further compression induces an abrupt shift in the PL peak position at 2.84 GPa for the $MAPbBr_3$-DMF:DMSO film and at 3.2 GPa for the $MAPbBr_3$-MAAc film, signaling another pressure-induced phase transition ($P_{tr}$-$2$). Corresponding optical microscopy images (**Figures 1g and 1h**) visually track these changes: both films deepen in color with increasing pressure, correlating with the red-shift and intensification of band-edge emission. As the PL weakens under higher compression, the films gradually fade to a pale yellow-green and eventually become nearly transparent upon complete PL quenching.

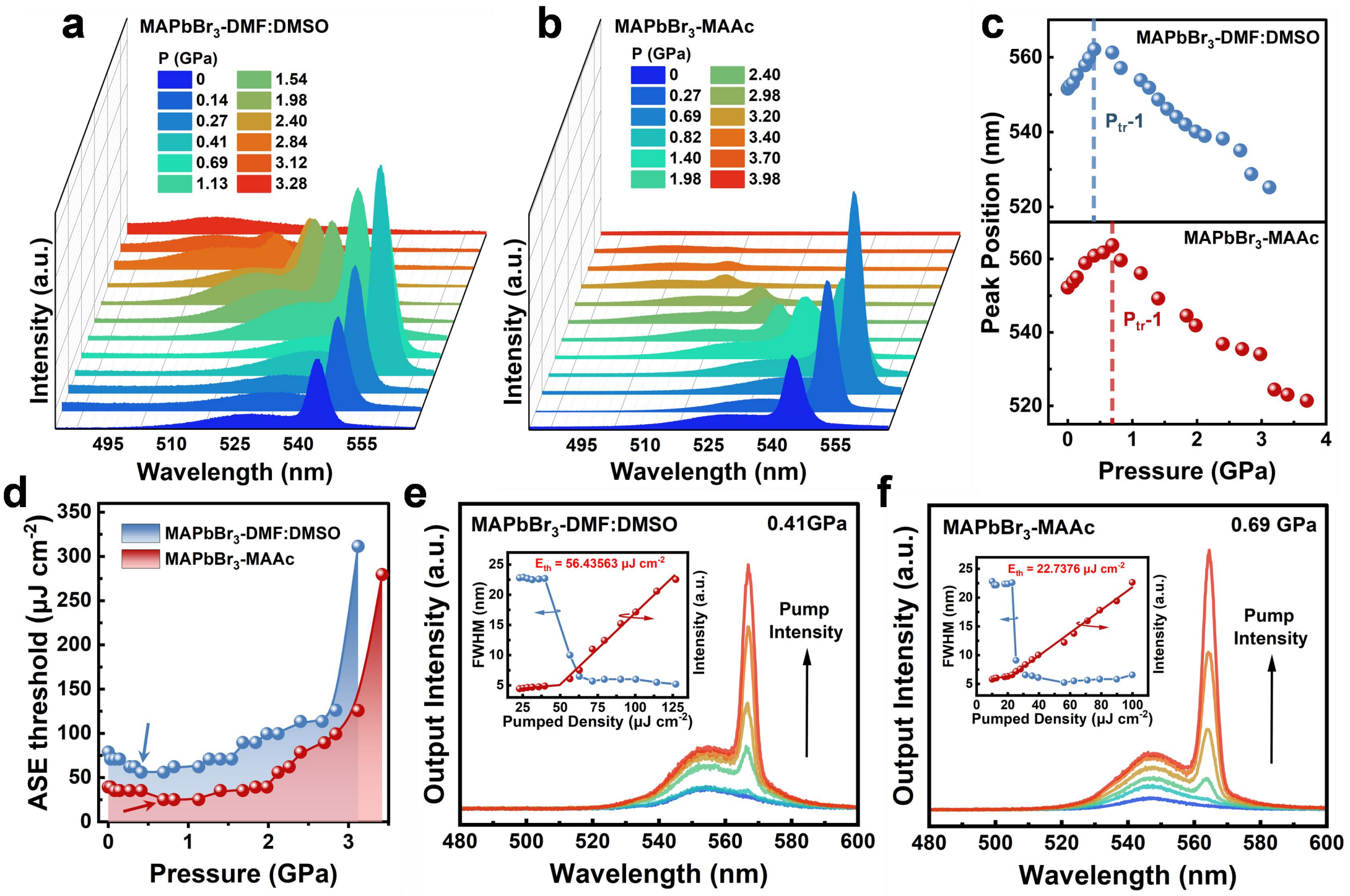

**Figure 2. Pressure-modulated ASE in $MAPbBr_3$ films.** Emission spectra of the (a) $MAPbBr_3$-DMF:DMSO and (b) $MAPbBr_3$-MAAc films at selected pressures under a fixed pump fluence of 790 μJ $cm^{-2}$. (c) Pressure dependence of the ASE emission peak wavelength for both films. (d) Evolution of the ASE threshold with pressure for both films. Emission spectra under varying pump fluences for the (e) $MAPbBr_3$-DMF:DMSO and (f) $MAPbBr_3$-MAAc films at ambient pressure. Insets: Corresponding plots of ASE threshold and FWHM as functions of pump fluence.

ASE represents a critical intermediate regime between spontaneous emission (SE) and coherent laser oscillation, exhibiting key laser-like attributes such as spectral narrowing and nonlinear intensity growth [30-32]. As such, the study of ASE provides fundamental insight into the optical gain properties and lasing potential of a material. Here, the pressure-dependent ASE characteristics of $MAPbBr_3$-DMF:DMSO and $MAPbBr_3$-MAAc films were investigated using a 355 nm nanosecond pulsed laser integrated within a custom-built *in situ.* high-pressure optical system (**Figure 1b**). Representative ASE spectra at selected pressures are shown in **Figures 2a** and **2b** (full dataset in **Figure S3**). The ASE intensity for both films exhibits a non-monotonic pressure dependence, initially increasing before subsequently weakening. The intensity maxima occur at 0.41 GPa for the $MAPbBr_3$-DMF:DMSO film and at 0.69 GPa for the $MAPbBr_3$-MAAc film, with the latter achieving an ASE intensity approximately three times greater at its optimum pressure ($P_{tr}$-$1$). As shown in **Figure**

**2c**, the ASE peak positions undergo pressure-induced spectral shifts consistent with the PL trends, featuring an initial red-shift followed by a blue-shift under higher compression.

At ambient pressure, as shown in **Figure S4**, the $MAPbBr_3$-MAAc film demonstrates a significantly lower ASE threshold (35 μJ $cm^{-2}$) compared to the $MAPbBr_3$-DMF:DMSO film (79 μJ $cm^{-2}$). This superior performance is attributed to the smooth, dense, and well-crystallized morphology enabled by the ionic liquid MAAc, which promotes higher optical gain and more efficient waveguiding [33]. The pressure evolution of the ASE threshold is summarized in **Figure 2d**. For the $MAPbBr_3$-DMF:DMSO film, the threshold decreases to a minimum of 56 μJ $cm^{-2}$ at 0.41 GPa, corresponding to a 29% reduction, before increasing until ASE quenching at 3.28 GPa. In striking contrast, the $MAPbBr_3$-MAAc film achieves a markedly lower minimum threshold of 22 μJ $cm^{-2}$ at 0.69 GPa-a 37% optimization-and sustains ASE emission up to a significantly higher quenching pressure of 3.98 GPa, underscoring its enhanced pressure robustness.

The transition from broad SE to sharp ASE is illustrated in **Figures 2e** and **2f**, which display emission spectra under varying pump fluences at the respective optimal pressures. Below threshold, only broad PL is observed. Upon exceeding the threshold, a narrow ASE peak emerges on the long-wavelength shoulder, with FWHM values of 4.9 nm and 4.6 nm for the $MAPbBr_3$-DMF:DMSO and $MAPbBr_3$-MAAc films, respectively, approximately 4 times narrower than their corresponding PL peaks. The insets clearly show the characteristic threshold behavior in both output intensity and spectral narrowing. That is, the $MAPbBr_3$-MAAc film consistently exhibits superior ASE performance under pressure, including a lower optimized threshold, higher emission intensity, and extended operational pressure range. This demonstrates that solvent engineering with MAAc not only enhances ambient performance but also significantly improves the functional resilience of perovskite gain media under extreme mechanical stress.

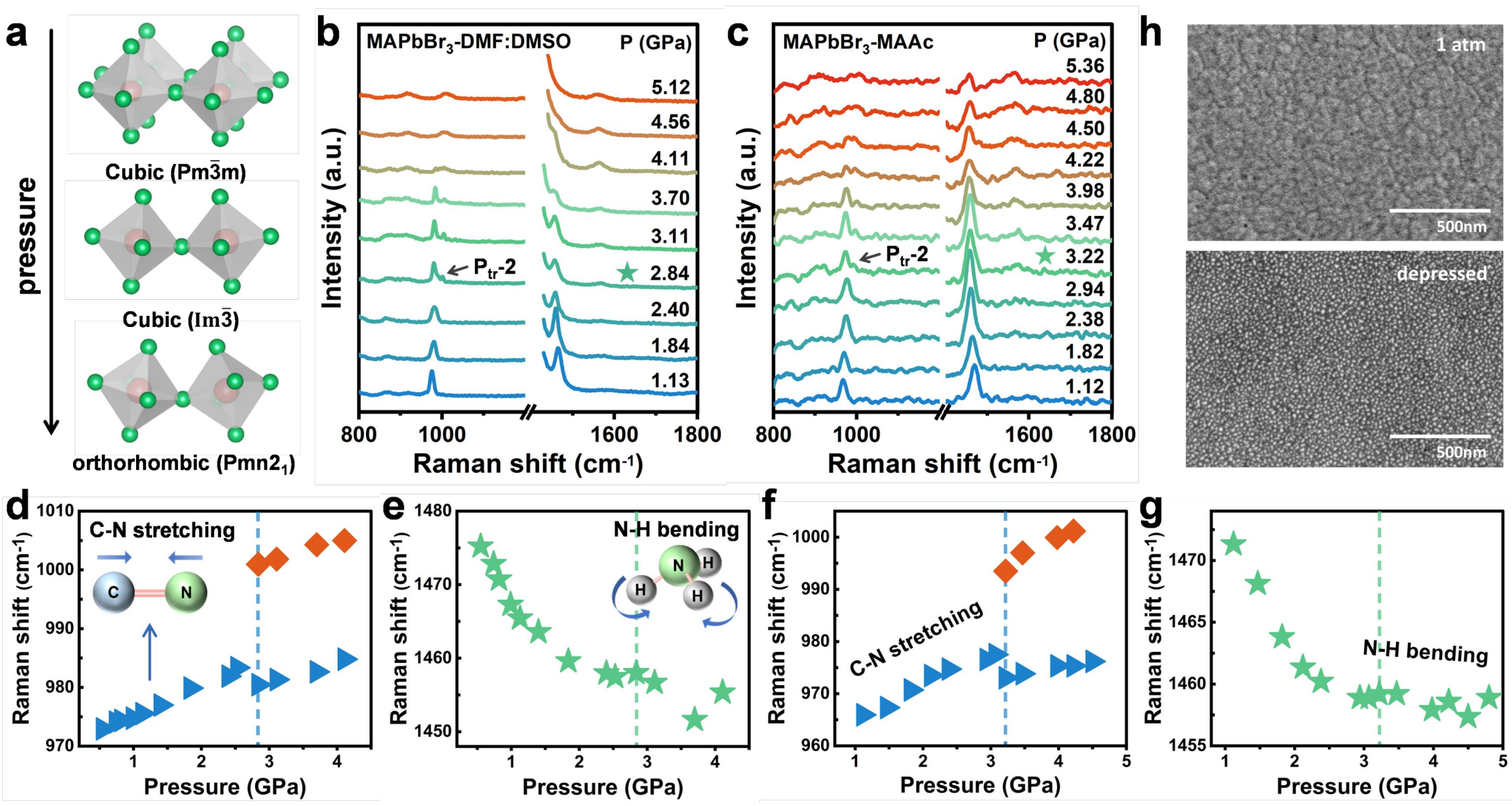

**Figure 3. Structural evolution and stability of $MAPbBr_3$ films under pressure.** (a) Schematic crystal structures of $MAPbBr_3$ across its high-pressure phases. High-pressure Raman spectra of the (b) $MAPbBr_3$-DMF:DMSO and (c) $MAPbBr_3$-MAAc films. Pressure dependence of the frequencies for two characteristic Raman modes in the (d, e) $MAPbBr_3$-DMF:DMSO and (f, g) $MAPbBr_3$-MAAc films. (h) Scanning electron microscopy (SEM) images of a freshly synthesized $MAPbBr_3$-MAAc film (up) and after one full compression-decompression cycle (down).

Given the intimate link between optical emission and crystal structure, *in situ.* high-pressure Raman spectroscopy was employed to elucidate the structural origins of the differing pressure robustness between the two film types. Under pressure, the $MAPbBr_3$ undergoes two phase transitions: $Pm\overline{3}m \rightarrow Im\overline{3} \rightarrow Pmn2_1$ (**Figure 3a**). The first phase transition aligns with previous research results, while the second one exhibits certain discrepancies, thus the measurements were initiated above 1 GPa. The Raman spectra of $MAPbBr_3$ (**Figures 3b, 3c**) feature low-frequency modes (<350 $cm^{-1}$) from Pb-Br vibrations and high-frequency modes (900-3000 $cm^{-1}$) from organic cations[34]. We focus on two prominent high-frequency peaks: the C-N stretching mode (~970 $cm^{-1}$) and the N-H bending mode (~1475 $cm^{-1}$)[35]. The pressure-induced frequency shifts of these modes are illustrated in **Figures 3d-g**. For the $MAPbBr_3$-DMF:DMSO film, the initial cubic phase ($Pm\overline{3}m$) persists until ~0.49 GPa ($P_{tr}$-*1*), followed by a transition to the tilted-cubic $Im\overline{3}$ phase, and transforms into the orthorhombic $Pmn2_1$ phase at 2.84 GPa ($P_{tr}$-*2*). The $MAPbBr_3$-MAAc film undergoes the same sequence of transitions but at systematically higher pressures, indicating delayed structural destabilization.

The C-N stretching mode exhibits a gradual blue-shift with pressure due to bond shortening. An abrupt red-shift and peak splitting at 2.84 GPa (DMF:DMSO) and 3.22 GPa (MAAc) mark the onset of the $Pmn2_1$ phase, consistent with the discontinuities observed in PL and ASE. Conversely, the N-H bending mode red-shifts under compression, likely due to strengthened hydrogen bonding that restricts molecular motion[36]. Its anomalous shifts near the phase transition may reflect metastable reorientations of the $MA^+$ cations. At pressures exceeding ~4.1 GPa, Raman peaks broaden significantly, signaling pressure-induced amorphization. Post-compression SEM imaging (**Figure 3h**) reveals that the initially continuous $MAPbBr_3$-MAAc film develops a granular morphology after a single compression-decompression cycle, confirming that high pressure induces irreversible microstructural changes, including likely grain fragmentation or defect generation. These structural insights directly explain the superior optical resilience of the MAAc-processed film. The delayed phase transitions and higher amorphization pressure observed in Raman spectra correlate with its extended operational window for both PL and ASE, demonstrating that solvent engineering enhances not only film morphology but also the intrinsic structural stability of the perovskite lattice under extreme pressure.

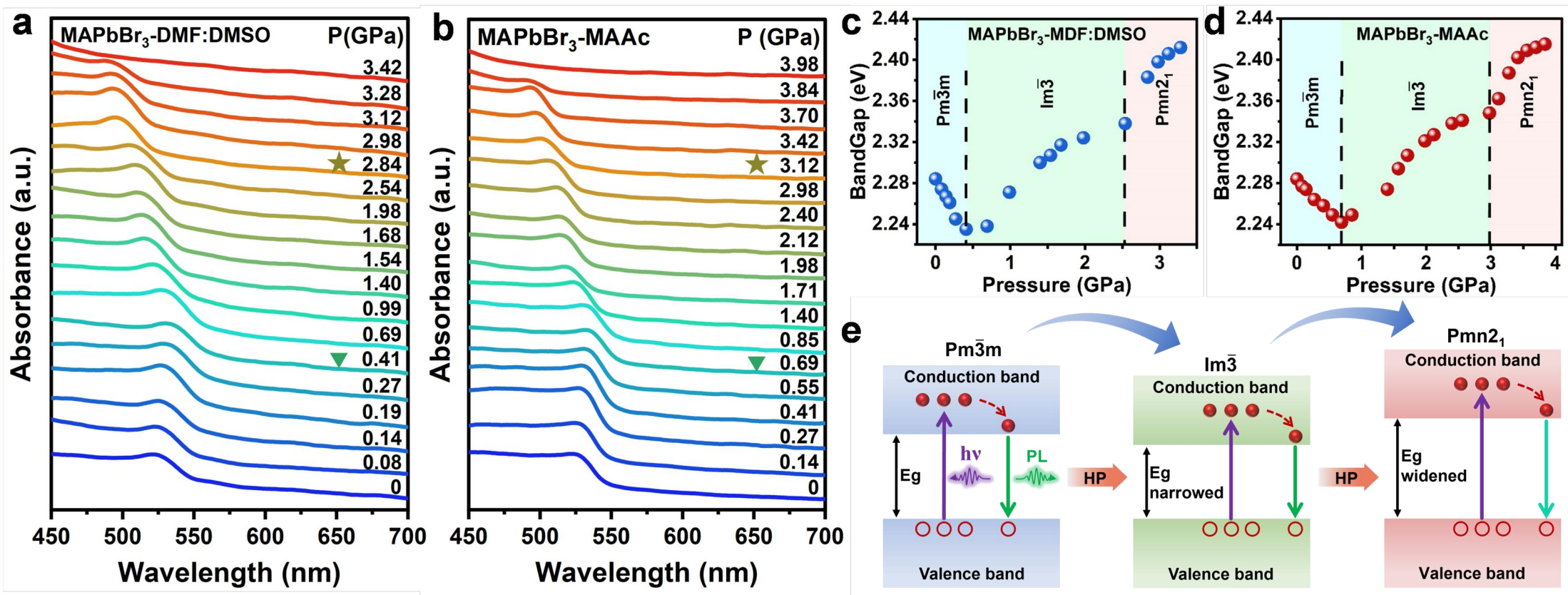

**Figure 4. Pressure-dependent optical absorption and bandgap evolution.** *In situ.* optical absorption spectra of the (a) $MAPbBr_3$-DMF:DMSO and (b) $MAPbBr_3$-MAAc films under pressure. Corresponding Tauc plots and derived direct bandgap energies as a function of pressure for the (c) $MAPbBr_3$-DMF:DMSO and (d) $MAPbBr_3$-MAAc films. (e) Schematic illustration of the proposed fluorescence emission mechanism for $MAPbBr_3$ films across different structural phases.

Pressure-induced structural phase transitions are intrinsically linked to modifications in the electronic band structure. To investigate this relationship, *in situ.* high-pressure steady-state absorption measurements were conducted on $MAPbBr_3$ films (**Figures 4a, 4b**). At ambient pressure, both films exhibit a well-defined absorption onset near 530 nm (**Figure S5**). With increasing pressure, the absorption edge shifts first to longer wavelengths (red-shift) and subsequently to shorter wavelengths (blue-shift). The corresponding direct bandgap energies were derived from Tauc plots (**Figures 4c, 4d**). Initially, both films possess a similar bandgap of ~2.28 eV. Upon compression within the cubic $Pm\bar{3}m$ phase, the bandgap narrows, reaching minima of 2.20 eV at 0.49 GPa for the $MAPbBr_3$-DMF:DMSO film and 2.19 eV at 0.69 GPa for the $MAPbBr_3$-MAAc film. These minima coincide with the onset of the first phase transition. Following the transition to the tilted-cubic $Im\bar{3}$ phase, the bandgap expands monotonically. A distinct discontinuity in this trend is observed at 2.84 GPa and 3.12 GPa for the $MAPbBr_3$-DMF:DMSO and $MAPbBr_3$-MAAc films, respectively, marking the second structural transition into the orthorhombic $Pmn2_1$ phase. Within this phase, the bandgap continues to increase, reaching approximately 2.41 eV before the absorption signal vanishes completely at 3.42 GPa and 3.98 GPa for the respective films. This bandgap evolution trajectory is fully consistent with the spectral shifts observed in both PL and ASE, while the systematically higher transition pressures for the $MAPbBr_3$-MAAc film confirm its enhanced structural resilience.

The underlying mechanism governing these optical changes is schematically summarized in **Figure 4e**. Theoretical studies indicate that pressure-induced bandgap modulation primarily arises from shifts in the conduction band minimum (CBM), composed of anti-bonding Pb *p* states highly sensitive to Pb-Br bond length[37]. In contrast, the valence band maximum (VBM), derived from Br *p* orbitals, remains relatively stable. At low pressures, lattice contraction strengthens orbital coupling, lowering the CBM energy and reducing the bandgap. Upon further compression beyond the first phase transition, pronounced octahedral tilting (Pb-Br-Pb angle ~165°[38]) weakens the Pb-Br orbital interaction, driving the CBM to higher energies and consequently widening the bandgap.

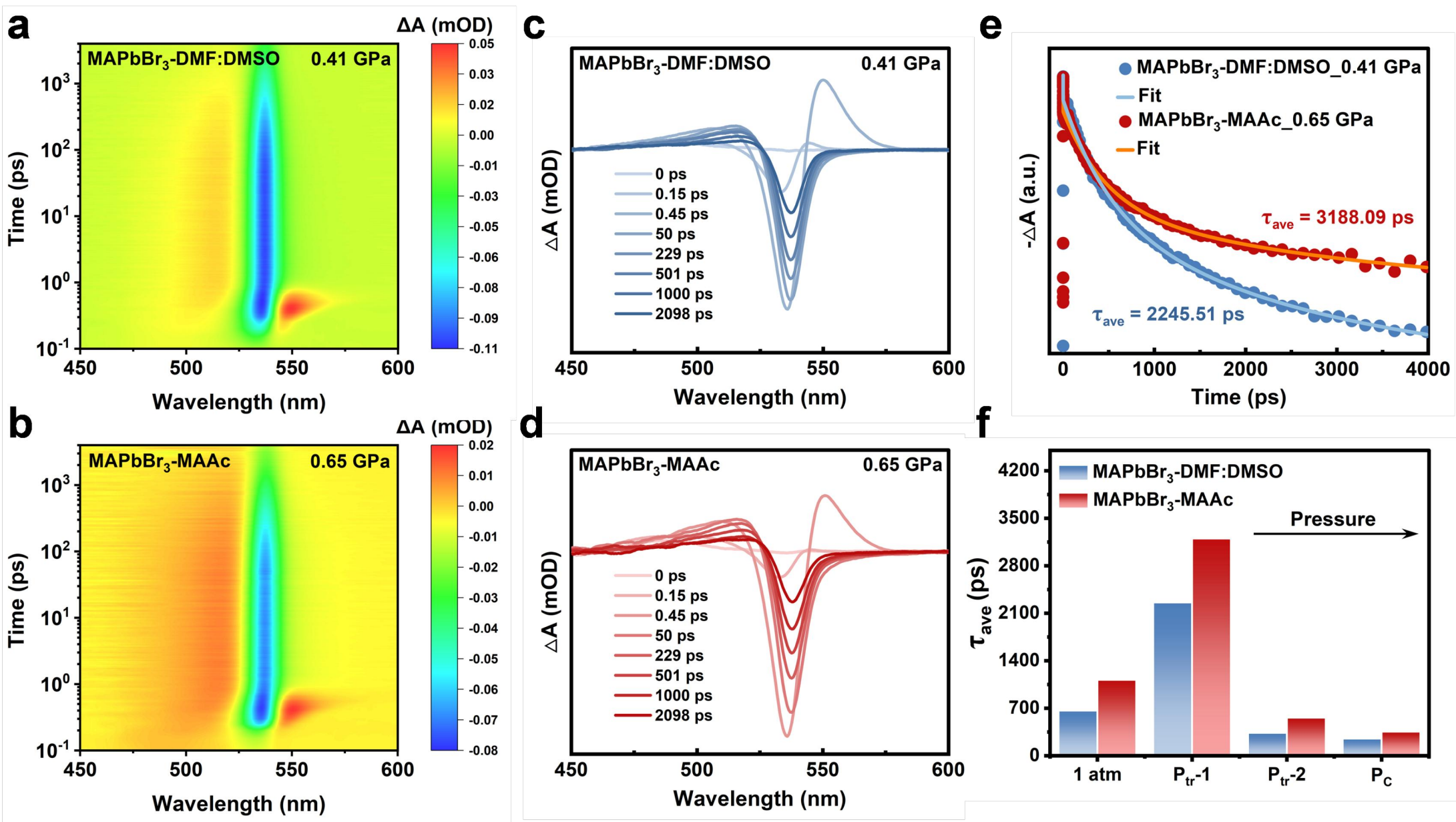

**Figure 5. Carrier dynamics of $MAPbBr_3$ films under pressure probed by TA spectroscopy.** Two-dimensional TA spectra (differential absorbance, ΔA) for the (a) $MAPbBr_3$-DMF:DMSO film at 0.41 GPa and (b) $MAPbBr_3$-MAAc film at 0.65 GPa. Selected TA spectra at delay times between 450 and 600 ps for the (c) $MAPbBr_3$-DMF:DMSO and (d) $MAPbBr_3$-MAAc films, extracted from (a) and (b), respectively. (e) Normalized kinetics of the band-edge PB signal for both films at their respective pressures of maximum PL intensity. (f) Average carrier lifetime ($\tau_{ave}$) for both films at key pressure points: the first phase transition ($P_{tr}$-1), the second phase transition ($P_{tr}$-2), and the critical quenching pressure ($P_C$).

To gain deeper insight into the origin of the performance disparity between the two film types, *in situ.* TA spectroscopy was employed to investigate their photocarrier dynamics under pressure. **Figures 5a-d** present the TA spectra and corresponding kinetics for the $MAPbBr_3$-DMF:DMSO film at 0.41 GPa and the $MAPbBr_3$-MAAc film at 0.65 GPa (their respective points of optimal PL/ASE intensity). A prominent photobleaching (PB) signal is observed at ~535 nm, aligning perfectly with the steady-state absorption edge. The PB decay kinetics were fitted with a tri-exponential function, revealing three distinct components: a fast decay ($\tau_1$, hot-carrier cooling), an intermediate decay ($\tau_2$, trap-assisted recombination), and a slow decay ($\tau_3$, bimolecular radiative recombination)[39-41]. The average carrier lifetime ($\tau_{ave}$) was calculated as:

$$\tau_{\text{ave}} = \frac{\sum_{i=1}^{n} A_i \tau_i^2}{\sum_{i=1}^{n} A_i \tau_i}$$

where $A_i$ and $\tau_i$ represent the amplitude and lifetime of each decay component, respectively. A striking difference is immediately apparent: at its optimal pressure, the $MAPbBr_3$-MAAc film exhibits a substantially longer $\tau_{ave}$ of 3188 ps, compared to 2246 ps for the $MAPbBr_3$-DMF:DMSO film (**Figure 5e**). Detailed fitting parameters across all pressures are provided in **Figure S6**, **Table S1**, and **Figure S7**. The pressure dependence of $\tau_{ave}$ (**Figure 5f**) mirrors the trend observed in PL and ASE performance, initially increasing to a maximum before decreasing upon further compression. Crucially, at every equivalent pressure, the $\tau_{ave}$ of the $MAPbBr_3$-MAAc film consistently surpasses that of its DMF:DMSO counterpart. This prolonged carrier lifetime, governed primarily by larger $\tau_2$ and $\tau_3$ components, directly indicates a lower trap density and more efficient radiative recombination pathways in the MAAc-processed film. These superior carrier dynamics provide a fundamental explanation for its enhanced ASE performance, including lower thresholds and higher emission intensities, under high-pressure conditions.

## 4. Conclusion

In summary, this work presents a comprehensive investigation into the high-pressure optical response and phase stability of $MAPbBr_3$ thin films engineered through distinct solvent strategies. The high-quality films prepared using MAAc ionic liquid demonstrate markedly superior pressure robustness compared to those fabricated using conventional DMF:DMSO solvents. This is evidenced by a systematic delay in all critical pressure thresholds: the MAAc-based films undergo phase transitions at higher pressures ($P_1$ and $P_2$) and sustain both PL and ASE emission up to 3.98 GPa, significantly exceeding the 3.34 GPa limit of the control films ($P_3$). Furthermore, across the entire investigated pressure range, the $MAPbBr_3$-MAAc films consistently exhibit lower ASE thresholds and longer carrier lifetimes, indicating more efficient radiative recombination and reduced defect-mediated losses under pressure. These results unequivocally demonstrate that solvent engineering with MAAc not only enhances the ambient performance of perovskite gain media but fundamentally strengthens their structural and functional resilience under extreme mechanical compression. This study therefore establishes solvent selection as a critical and viable strategy for tailoring the high-pressure stability of metal halide perovskites, providing a valuable design principle for developing durable optoelectronic and laser devices capable of operating in demanding environments.